\documentclass[aps,prl,amsmath,amssymb,reprint,
longbibliography,groupedaddress]{revtex4-2}
\usepackage[utf8]{inputenc}
\usepackage[hidelinks]{hyperref} 
\usepackage{graphicx, color}
\usepackage{mathtools}
\newcommand{\red}[1]{{\color{red}#1}}

\usepackage{mycommands}
\usepackage{bm}

\begin{document}

\title[SC-TST]{Semiclassical Spin Exchange
via Temperature-Dependent Transition States}
\author{Debaarjun Mukherjee}
\author{Jeremy O. Richardson}
\email[Correspondence email address: ]{jeremy.richardson@phys.chem.ethz.ch}
\affiliation{Institute of Molecular Physical Science, ETH Zürich, 8093 Zürich, Switzerland} 

\date{\today}

\begin{abstract}
Spin-exchange collisions have been widely studied in recent years, and various quantum-mechanical scattering approaches have been developed to calculate the rates. However, these methods based on global knowledge of wavefunctions can be computationally demanding and do not offer a simple mechanistic interpretation. Here, we present a new semiclassical transition-state theory (SCTST) derived from first principles to describe the nonadiabatic transition between two states which differ only in their spins, where classical TST and Landau--Zener theory fail.
We apply our theory to describe the spin-exchange collision between the nuclear spin of $\prescript{3}{}{\text{He}}$ and the electronic spin of $\prescript{23}{}{\text{Na}}$.
SCTST reveals that the reaction proceeds via a temperature-dependent transition state, determined by an intricate compromise between minimizing the activation energy and maximizing the hyperfine coupling.
It further demonstrates the importance of quantum delocalization effects prevalent in spin exchange even when tunneling is suppressed and successfully explains the weak temperature dependence of the rate. Moreover, since it depends only on local information at a single point, the computational cost is significantly reduced.

\end{abstract}

\maketitle
Utilizing the spin degree of freedom for information storage and fabrication of quantum materials has been one of the key goals in the field of spintronics \cite{Spintronics_review}. The bottleneck for this process is the generation and control of the spin without inducing any change in the charge of the system \cite{spintronic_npj_q}.
The nuclear spin of odd isotopes of noble gases has a non-zero quantum number and a long coherence time \cite{RevModPhys_Walker_review}, as it is well shielded by the electron cloud from external agents. However, these nuclear spins can be coherently controlled by coupling them with the electronic spins of alkali metals via spin-exchange processes \cite{coherent_coupling,katz2015coherent_coupling}, making them important candidates for constructing quantum materials \cite{PRXQuantum_Firstenberg}. Other forms of spin exchange 
are prevalent in atom--photon collisions \cite{yang_atom_photon_spinflip} and recently shown to cause energy transfer between atomic hyperfine and molecular rotational states \cite{hyperfine_to_rotational}. Thus, it is crucial to understand the mechanism of the spin-exchange process in order to improve the rational design of quantum materials. 
 
Various quantum scattering methods have been proposed over the years to calculate the rate constants for spin-exchange processes \cite{walker_theory,Romalis_theory,main_tsherbul,walker1989estimates,walter_theory,anisotropic_tsherbul}. For instance, the exchange between the nuclear spin of $^3$He and the electronic spin of the valence electron of $^{23}$Na is known to be mediated by short-range hyperfine interactions \cite{Ramsey_main}, of which the isotropic Fermi contact interaction plays the dominant role \cite{walker_theory}. 
After removing the centre-of-mass motion, the Hamiltonian for the spin-exchange process in the absence of a magnetic field is \cite{walker_happer_rev}
 \begin{equation} \label{H}
     \hat{H}=-\frac{\hbar^2}{2m}\frac{\partial^2}{\partial r^2} + \frac{\hat{L}^2}{2mr^2}+V(r)+ A(r)\hat{I}\cdot\hat{S} ,
 \end{equation}
 where $m$ is the reduced mass, $r$ is the internuclear distance, $\hat{L}$ is the spatial angular-momentum operator and $V(r)$ is the potential energy.
The final term couples the nuclear and electronic spin angular-momentum operators, $\hat{I}$ and $\hat{S}$, via the Fermi contact interaction,
 \begin{equation} \label{Afc_equation}
    A(r)=\frac{16\pi}{3}\frac{\mu_0\mu_\mathrm{He}}{I}\rho_{\mathrm{He}}(r) .
\end{equation}
Here, $\mu_0$ and $\mu_\mathrm{He}$ are the Bohr magneton and nuclear magnetic moment of $^3$He, $I=\frac{1}{2}$ is the nuclear spin quantum number and $\rho_\mathrm{He}(r)$ is the electron spin density at the nucleus of He. We neglect the long-range anisotropic spin-dipolar part of the hyperfine coupling, which only becomes important at very low temperatures
\cite{anisotropic_tsherbul}.
  
Additionally, since we are describing spin exchange and not spin relaxation, the spin--rotation term, which couples the electronic spin with the spatial angular momentum (and is proportional to Na's weak spin--orbit coupling \cite{walker_happer_rev}) has been omitted in the expression.

Although quantum scattering methods can in principle solve this problem numerically exactly, they involve tedious calculations of the partial waves and do not provide an intuitive understanding of the spin-exchange mechanism. Furthermore, they require global knowledge of the system.  However, calculating the potential as well as the hyperfine coupling between the atoms at large distances can be difficult because of the effects of various long-range interactions.

On the other hand, classical methods such as Marcus theory \cite{Marcus1993review} and nonadiabatic transition-state theory (TST) \cite{Lorquet1988NATST,Lykhin2016NATST} utilize Landau--Zener transmission coefficients \cite{Landau1932LZ,Zener1932LZ} based on information at the minimum-energy crossing point (MECP) of the reactant and product potential energy surface (PES) and have been used to calculate spin-flip rates (also known as intersystem crossing) in molecules in an elegant manner \cite{Harvey2014review}. However, in the case of spin-exchange collisions, the reactant and product PESs are identical and thus ``cross'' at every point, 
which causes the classical rates to diverge.
Moreover, due to the absence of a unique transition state, the mechanism is unclear. 
Hence, it is necessary to devise an improved transition-state theory which can describe the spin-exchange mechanism.

In this Letter, we describe a new semiclassical transition-state theory (SCTST) which stems from instanton theory \cite{Miller1975semiclassical, Miller1990SCTST,Coleman1977ImF} and its generalization to nonadiabatic reactions \cite{inverted,GRperspective}. 
It uses path integrals rather than wavefunctions and thus offers a different way of looking at the phenomenon. This method requires only local knowledge of the system at a single geometry, called the ``hopping point'', which is defined as the minimum of the \emph{temperature-dependent effective potential}, thus significantly reducing the computational cost. Nonetheless, it is able to predict the spin-exchange rates with good accuracy, which confirms that the hopping point is the key to understanding the mechanism.
.

We take the initial (reactant) state of the system to be $\ket{\mathrm{R}}=\ket{\uparrow\downarrow}$ and the final (product) state to be $\ket{\mathrm{P}}=\ket{\downarrow\uparrow}$, where the two arrows represent the nuclear and electronic spins.
The Hamiltonians for each state are thus $\hat{H}_n=\braket{n|\hat{H}|n}$ for $n \in \{\mathrm{R},\mathrm{P}\}$, which have kinetic energy identical to the first two terms of \eqn{H} and an effective potential energy $V_{n}(r)=V(r)+A(r)/4\approx V(r)$.
The two states are coupled by
\begin{equation}
    \Delta(r) = \bra{\uparrow\downarrow}A(r)\hat{I}\cdot\hat{S}\ket{\downarrow\uparrow} 
      =\frac{A(r)}{2}. 
\end{equation}

The thermal rate is given by
\begin{equation} \label{FGR}
k \, Z_\mathrm{R} = \frac{1}{2\pi\hbar} \int_0^\infty N(E) \,\text{e}^{-\beta E} \, \mathrm{d} E,
\end{equation}
where the cumulative reaction probability (assuming the coupling to be weak) is defined using Fermi's golden rule \cite{Dirac1927radiation}
\begin{multline}
N(E_\mathrm{R}) = (2\pi)^2 \sum_\ell (2\ell+1) \int|\Delta^\ell(E_\mathrm{R},E_\mathrm{P})|^2 \\
\times\delta(E_\mathrm{R} - E_\mathrm{P})\, \mathrm{d}E_\mathrm{P}.
\end{multline}
Here, the scattering states (with angular momentum quantum number $\ell$ and collision energy $E_n$) obey $\hat{H}_{n}\psi^{\ell}_n(E_n)=E_n\psi^{\ell}_n(E_n)$ 
 and are coupled by
 $\Delta^{\ell}(E_\mathrm{R},E_\mathrm{P})=\bra{\psi_\mathrm{R}^\ell(E_\mathrm{R})}\Delta\ket{\psi_\mathrm{P}^\ell(E_\mathrm{P})}$.
 Finally, $Z_\mathrm{R}=(m/2\pi\beta\hbar^2)^{3/2}$
is the reactant partition function.

As we wish to avoid computing scattering wavefunctions, we now present an alternative formula in a basis-independent representation obtained using the Fourier transform of the delta function \cite{Wolynes1987nonadiabatic,GRperspective}:
\begin{multline}
    k \, Z_\mathrm{R}=\frac{1}{\hbar^2} \int_{-\infty}^{\infty}\Tr\left[\hat{\Delta} \,\eu{-(\beta\hbar-\tau-\iu t) \hat{H}_\mathrm{R}} \hat{\Delta} \,\eu{-(\tau+\iu t)\hat{H}_\mathrm{P}} \right] \rmd t,
\end{multline}
where we have introduced an arbitrary component of imaginary-time, $\tau$.
The most direct way to obtain a classical rate equation is by evaluating the trace as a phase-space integral \cite{ChandlerET},
\begin{multline}\label{phase_space}
    k_\mathrm{cl}\,Z_\mathrm{R}=\frac{1}{\hbar^2}\frac{1}{(2\pi\hbar)^3}\iiint_{-\infty}^{\infty}\Delta^2(\mathbf{x}) \, \eu{-\beta ||\mathbf{p}||^2/2m}\\
    \times\eu{-\beta V_{\mathrm{R}}-(V_\mathrm{P}-V_\mathrm{R})(\tau+\mathrm{i}t)/\hbar}\, \mathrm{d}t\,\mathrm{d}\mathbf{p}\,\mathrm{d}\mathbf{x},
\end{multline}
with $\mathbf{x}$ being the internuclear vector and $\mathbf{p}$ its conjugate momentum.
In the general case, the time integral gives $2\pi\delta[V_\mathrm{R}(\mathbf{x})-V_\mathrm{P}(\mathbf{x})]$ and thus forces the hop to occur at a crossing point, as in Marcus theory.
However, it becomes immediately evident that for completely overlapping potentials where $V_\mathrm{R}(\mathbf{x})=V_\mathrm{P}(\mathbf{x})$ for all $\mathbf{x}$, which is the case for spin-exchange collisions, the time integral becomes undefined and the classical approach is no longer valid.

To solve this problem, 
we return to quantum mechanics
and express the trace in the path-integral formalism. First, we consider a general case where $V_\mathrm{R}$ and $V_\mathrm{P}$ are different functions before studying the implications of setting them to be identical. Writing \eqn{FGR} in terms of quantum propagators $K_n$, 
the rate expression becomes \cite{Wolynes1987nonadiabatic,InstReview},
\begin{multline}
    k \, Z_\mathrm{R}
    =\frac{1}{\hbar^2} \iiint \Delta(\mathbf{x}')K_\mathrm{R}(\mathbf{x}',\mathbf{x}'',\beta\hbar-\tau-\iu t) \\
     \times\Delta(\mathbf{x}'')K_\mathrm{P}(\mathbf{x}'',\mathbf{x}',\tau+\iu t)
    \, \rmd \mathbf{x}' \rmd \mathbf{x}'' \rmd t .
\end{multline}

To obtain a high-temperature limit for the rate, we 
introduce the semiclassical propagators \cite{GutzwillerBook,Miller1971density,InstReview} in the short-time limit:
$K_n(\mathbf{x}',\mathbf{x}'',\tau_n) \simeq \left({m}/{2\pi\hbar\tau_n}\right)^{3/2} \eu{-S_n/\hbar}$,
with the classical action \cite{Feynman,Kleinert,GoldenGreens}:
\begin{equation}
    {S}_n(\mathbf{x}',\mathbf{x}'',\tau_n)\simeq\frac{m}{2\tau_n}||\mathbf{x}_-||^2 +\tau_nV_{n}({\mathbf{x}_+}).
\end{equation}
For simplicity, we have transformed the vectorial variables to $\mathbf{x}_+={(\mathbf{x}'+\mathbf{x}'')}/{2}$ and $\mathbf{x}_{-}={(\mathbf{x}'-\mathbf{x}'')}$ with radial components $r_+$ and $r_-$ respectively.

In many golden-rule processes \cite{GRperspective}, one can treat $\Delta(\mathbf{x})$ as slowly varying; however, this so-called Condon approximation is no longer valid here as it leads to divergent rates.
To extend the theory for a coupling which strongly depends on the internuclear position, we move the $\Delta(\mathbf{x})$ terms to the exponential \cite{CIinst} and perform a steepest-descent integration
around the stationary point of the effective action $S_\mathrm{eff}=S_\mathrm{R}+S_\mathrm{P}- \hbar\ln\frac{\Delta(\mathbf{x}')\Delta(\mathbf{x}'')}{\Delta_0^2}$.
This gives
\begin{equation}\label{Instanton Theory}
    k_\mathrm{cl} \, Z_\mathrm{R} = \sqrt{2\pi\hbar} \, \frac{\Delta_0^2}{\hbar^2}Z_\mathrm{rot} \sqrt{\frac{m^2}{\tau_\mathrm{R}\tau_\mathrm{P}}} \sqrt{\frac{1}{-\Sigma_\mathrm{eff}}} \, \eu{-S_\mathrm{eff}/\hbar} ,
\end{equation}
where $Z_\mathrm{rot}$ is the standard rotational partition function measured at the hopping point. 
Additionally, $\tau_\mathrm{R}=\beta\hbar-\tau$ and $\tau_\mathrm{P}=\tau$ correspond to the imaginary time spent on the
reactant and product states (which is determined by the stationary-action condition),
and an arbitrary constant $\Delta_0$ has been introduced to take care of the units.
The negative sign before $\Sigma_\mathrm{eff}$ comes from the Cauchy--Reimann relations which allow us to perform the derivatives in imaginary time:
\begin{align}
    \renewcommand*{\arraystretch}{1.5}
    \Sigma_\mathrm{eff} = \begin{vmatrix} 
    \pder{S_\mathrm{eff}}{{r}^2_{-}}{} & \pders{S_\mathrm{eff}}{{r}_{-}}{{r}_{+}} & \pders{S_\mathrm{eff}}{{r}_{-}}{\tau} \\
    \pders{S_\mathrm{eff}}{{r}_{+}}{{r}_{-}} & \pder{S_\mathrm{eff}}{{r}_{+}^2}{} & \pders{S_\mathrm{eff}}{{r}_{+}}{\tau} \\
    \pders{S_\mathrm{eff}}{\tau}{{r}_{-}} & \pders{S_\mathrm{eff}}{\tau}{{r}_{+}} & \pder{S_\mathrm{eff}}{\tau^2}{}
    \end{vmatrix}  .
\end{align}
\begin{figure}
\includegraphics[width=\columnwidth]{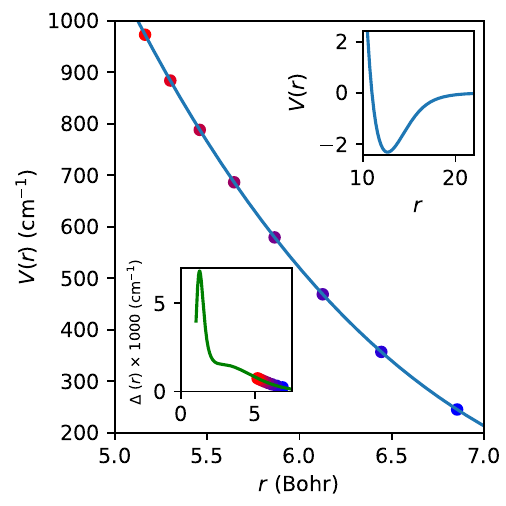 }
\caption{Hopping points, $r^\ddag$, from 200 to 900\,K in steps of 100\,K. With increasing temperature, the hopping point (also known as the transition state) climbs to higher energies to access larger coupling. The insets show the potential energy curve and the coupling along a wider range of internuclear distances.  The shallow minimum in the potential plays no role in the semiclassical theory.}
\label{potential}
\end{figure}

In the short-time limit,
\begin{align}
    \Sigma_\mathrm{eff}^\mathrm{cl}= 
    \begin{vmatrix}
        \frac{m}{\tau_\mathrm{R}}+\frac{m}{\tau_\mathrm{P}} - \frac{\Xi}{2}& 0 &{0} \\
        {0} & {\beta\hbar\Bar{\Omega}} - 2\Xi & \frac{1}{\sqrt{2}}(\kappa_\mathrm{R}-\kappa_\mathrm{P}) \\
       {0} & {\frac{1}{\sqrt{2}}(\kappa_\mathrm{R}-\kappa_\mathrm{P})} & 0
    \end{vmatrix},
\end{align}
with $\Bar{\Omega}=\frac{\kappa_\mathrm{R}\Omega_\mathrm{P}-\kappa_\mathrm{P}\Omega_\mathrm{R}}{\kappa_\mathrm{R}-\kappa_\mathrm{P}}$ 
and $\Xi= \frac{\partial^2 \ln\Delta}{\partial r^2}$, where $\kappa_{n}$ and $\Omega_{n}$ are the first and second derivatives of the potential at the hopping point $r^\ddag$. These expressions are similar to the ones derived in previous works \cite{GoldenGreens,PhilTransA} except for the additional $\Xi$ terms in the determinant. In fact, 
in the high-temperature limit, we recover a form similar to that of nonadiabatic TST,
\begin{equation} \label{HT_general}
    k_\mathrm{cl}=\sqrt{\frac{2\pi m}{\beta \hbar^2}}\frac{\Delta(r^\ddag)^2}{\hbar |\kappa_\mathrm{R}-\kappa_\mathrm{P}|}\frac{Z_\mathrm{rot}}{Z_\mathrm{R}} \, \eu{-\beta V^\ddag},
\end{equation}
This formula is related to Marcus theory, which is recovered from the additional assumption of harmonic potentials. 
One recognizes that the prefactor is proportional to the Landau--Zener factor in the weak-coupling limit. 


An interesting phenomenon arises when we attempt to apply this classical theory to spin-exchange collision. In this case, the reactant and product surfaces are identical everywhere, 
such that $\kappa_\mathrm{R}=\kappa_\mathrm{P}$, which causes the Landau--Zener factor and hence the classical rate to diverge.
However, in contrast to the phase-space approach of \eqn{phase_space}, it is now clear how the issue can be fixed.

The problem stems from the fact that the 
determinant $\Sigma_\mathrm{eff}^\mathrm{cl}$ is zero when $\kappa_\mathrm{R}=\kappa_\mathrm{P}$.
Thus, instead of truncating the action at the lowest order, we are forced to include some quantum-mechanical effects,
\begin{multline}\label{action_x}
    {S}_{n}(\mathbf{x}',\mathbf{x}'',\tau_{n})\simeq\frac{m}{2\tau_n}||\mathbf{x}_-||^2 +\tau_nV_{n}({\mathbf{x}_+})
    -\frac{\kappa_{n}^2\tau_{n}^3}{24m},
\end{multline}
where the final term is recognized from the action of a linear potential \cite{Feynman,PhilTransA} and accounts for quantum-mechanical delocalization.
In this way, SCTST goes beyond the classical rate theories by including certain quantum effects, which, as we now show, is crucial to describe the spin-exchange process.
The semiclassical determinant now becomes,
\begin{align}
    \renewcommand*{\arraystretch}{0.5}
    \Sigma_\mathrm{eff}^\mathrm{sc}\simeq \begin{vmatrix}
        \frac{4m}{\beta\hbar}& {0} & {0} \\
        {0} & {\beta\hbar\Omega} -2\Xi & {0} \\
       {0} & {0} & -\frac{\beta \hbar \kappa^2}{4m}
    \end{vmatrix},
\end{align}
where $\kappa=\kappa_\mathrm{R}=\kappa_\mathrm{P}$ and $\Omega=\Omega_\mathrm{R}=\Omega_\mathrm{P}$.
Here, we have also used the symmetry 
to determine $\tau_\mathrm{R}=\tau_\mathrm{P}={\beta\hbar}/{2}$
and have neglected the subdominant $\Xi$ compared to $\mathcal{O}(1/\beta)$.
Note that the final term in \eqn{action_x} is clearly subdominant in the high-temperature limit.
However, it is nonetheless essential to describe the quantum fluctuations correctly to obtain a non-zero determinant and hence a well-defined rate.

The semiclassical rate constant is therefore defined as
\begin{equation} \label{HT_spin-exchange}
    k_\mathrm{SCTST}=\sqrt{\frac{2\pi m}{\beta \hbar^2}}\frac{\Delta^2(r^\ddag)}{\hbar |\kappa/{2}|}\frac{Z_\mathrm{rot}}{Z_\mathrm{R}}\frac{1}{\beta\hbar\omega}\,\eu{-\beta V^\ddag},
\end{equation}
where $m\omega^2=\Omega-2\Xi/{\beta\hbar}$ and $r^\ddag$ is the minimum of the temperature-dependent effective potential, $V_\text{eff}(r)=V(r)-\frac{2}{\beta}\ln\frac{\Delta(r)}{\Delta_0}$. 
This rate expression for SCTST appears strikingly similar to \eqn{HT_general}.  However, it is marked by certain important differences, in particular that the prefactor differs from the Landau--Zener form and thus avoids the problem of divergence.

\begin{figure}
\includegraphics[width=\columnwidth]{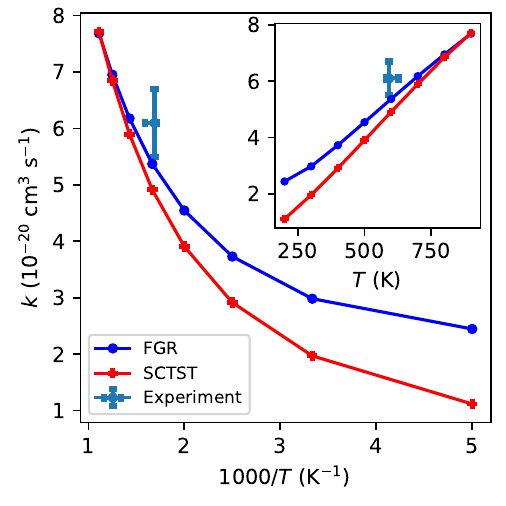}
\caption{Rate constants for the spin-exchange between $^3$He and $^{23}$Na compared with the experimental measurement of Ref.~\cite{borel_exp}. The semiclassical rates converge to the quantum-mechanical rate at high temperatures. The inset shows that the rate constants are only roughly linear with temperature. Experiments suggest that the rate is roughly constant (within error bars) over the small temperature range measured (as indicated), which is in line with our explanation of the rate constant only changing slightly with temperature.}
\label{fig:rates}
\end{figure}

In this way, SCTST provides an intuitive picture of the whole spin-exchange mechanism where the most probable approach distance is described by the hopping point, $r^\ddag$, also known as the transition state. 
There is a competition between the activation energy required to reach this point and the gain in increasing the coupling.
For this reason, the spin-exchange process does not follow the usual Arrhenius law, as increasing the temperature raises the activation energy in such a way that the rate constant only changes slightly.
This is in agreement with experiments \cite{borel_exp,Baranga_PRL}, where no temperature dependence of the spin-exchange rate was observed over the small temperature range measured.

In order to apply the SCTST method to the \ce{^3He + Na} spin-exchange reaction, we calculated the potential energy using CCSD(T)/cc-pVTZ and the Fermi contact interaction using density functional theory with the $\omega$B97X-D3 functional and the aug-PCSSEG-3 basis along with large auxillary basis sets included via the AUTOAUX option in the quantum-chemistry package {\sffamily ORCA} \cite{Neese2012orca} \footnote{Both curves were fitted to a univariate 5th-order spline to smoothen them. A smoothening factor of $10^{-11}$\,a.u was applied for $V(r)$ and $10^{-15}$\,a.u for $\Delta(r)$.}.

Figure~\ref{fig:rates} shows that the SCTST rates are in good agreement with the quantum rates calculated via Fermi's golden rule with scattering wavefunctions. 
SCTST also captures the qualitative trend of the rates decreasing roughly linearly with temperature as suggested previously by Walter et al.\ \cite{walter_theory} who had proposed an empirical temperature scaling relation for the rate constant. However, this apparent linear dependence of the rate constant on temperature is fortuitous and we see a deviation from linearity when the Fermi contact is replaced by a simpler exponential function (not shown).

Note that the strong deviation from the usual Arrhenius form is a consequence of the \emph{temperature-dependent transition states} and not an indication of tunneling effects.
In fact, the close agreement between our SCTST rate (which negelects tunneling) and FGR confirms that tunneling is not important in this case.
In particular, including the higher-order term from \eqn{action_x} into the exponent has a very minor effect on the result.

Our first-principles study also agrees with previous works \cite{walker1989estimates,main_tsherbul}, where the atomic model was used to approximate the Fermi contact with an empirical enhancement factor obtained by comparing theoretical calculations of the frequency-shift enhancement factor with experiment. The experimental rate constant reported by Borel et al.\ \cite{borel_exp} is $(6.1\pm 0.6)\times 10^{-20}\,\text{cm}^3\,\text{s}^{-1}$ at $319^\circ$C which is very close to our quantum rate of $5.3 \times 10^{-20}$\,$\text{cm}^3\,\text{s}^{-1}$ and our SCTST rate of $4.8 \times 10^{-20}$\,$\text{cm}^3\,\text{s}^{-1}$. This highlights that the error made by SCTST with respect to the quantum-mechanical rate is less than the uncertainty of the experimental result.

\begin{figure}
\includegraphics[width=\columnwidth]{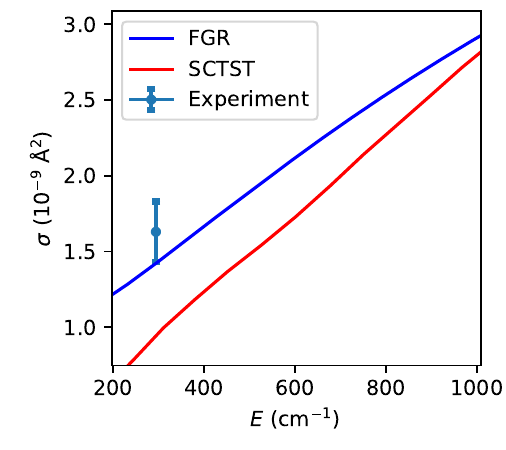}
\caption{Quantum and semiclassical spin-exchange cross-sections compared with the experimental measurement of Ref.~\cite{soboll1972spin}. Note that the experiment was carried out at 150\,K and we have plotted the reported cross-section at the corresponding thermal energy. The semiclassical spin-exchange results are obtained via the inverse
Laplace transform of the thermal rates while the quantum results are calculated from Fermi's golden rule [Eq.~\eqref{FGR}].}
\label{fig:sigma}
\end{figure}

In addition to calculating thermal rates, we can also predict reaction cross-sections using SCTST\@.
Formally, the cumulative reaction probabilities are related to the thermal rate by an inverse Laplace transform of Eq.~\eqref{FGR},
\begin{equation}
    N(E)=\frac{1}{2\pi \iu}\int_{\beta_\mathrm{sp}-\iu\infty}^{\beta_\mathrm{sp}+\iu\infty}\eu{\beta E + \ln F(\beta)}\, \mathrm{d}\beta,
\end{equation}
where $F(\beta)=k(\beta)\,Z_\mathrm{R}(\beta)$.
Making the stationary-phase approximation \cite{DosTMI,JoeFaraday} for a given energy, the stationary point $\beta_\mathrm{sp}$ satisfies $E=-\left(\frac{\mathrm{d}\ln F(\beta)}{\mathrm{d}\beta}\right)_{\beta=\beta_\mathrm{sp}}$,
with the cumulative reaction probability
\begin{equation}
    N(E)\simeq\left(2\pi\frac{\mathrm{d}^2\ln{F(\beta)}}{\mathrm{d}\beta^2}\right)^{-1/2}_{\beta=\beta_\mathrm{sp}}\eu{\beta_\mathrm{sp}E+\ln{F(\beta_\mathrm{sp})}}.
\end{equation}
Finally, the spin-exchange cross-section can be obtained as $\sigma(E)={\pi\hbar^2}{N(E)}/{2mE}$ \cite{Levine}. 
The results in Fig.~\ref{fig:sigma} show that at higher energies the semiclassical approximation approaches the quantum result and shows only a minor error at lower energies. Our calculations are also in good agreement with the spin-exchange cross-section measured by Soboll \cite{soboll1972spin}.

In conclusion, we have developed a new semiclassical transition-state theory for nonadiabatic transitions between two identical PESs and used it to explain the mechanism of the spin-exchange phenomenon in a simple way. This clarifies the role of quantum effects in the process---although there is negligible tunneling, it is necessary to include quantum delocalization in order to avoid divergences. The results are in good agreement with experiment and could in principle be systematically improved by increasing the accuracy of the electronic-structure calculations as well as by perturbatively incorporating extra effects such as anharmonicity and higher-order contributions beyond Fermi's golden rule \cite{lawrence2023perturbatively,trenins2022nonadiabatic}. SCTST can be extended further to include anisotropic and spin-orbit couplings \cite{anisotropic_tsherbul,spin_orbit_ion_atom} along with the effects of electric and magnetic fields \cite{electric_field_spin_flip,krems_magnetic_depolarization,katz_decoherence} by incorporating them in the electronic-structure calculations. Moreover, the method is not limited to atom--atom scattering but can also be used to study spin-exchange collisions between polyatomic molecules or atom--surface collisions. In this way, the new SCTST approach opens up possibilities for a better understanding of the spin-exchange process to aid the rational design of new quantum technologies by controlling the spin-polarization timescales.

The authors acknowledge financial support from the Swiss National Science Foundation through SNSF project titled ‘Nonadiabatic effects in chemical reactions’.

\bibliography{references,extra}

\begin{thebibliography}{52}%
\makeatletter
\providecommand \@ifxundefined [1]{%
 \@ifx{#1\undefined}
}%
\providecommand \@ifnum [1]{%
 \ifnum #1\expandafter \@firstoftwo
 \else \expandafter \@secondoftwo
 \fi
}%
\providecommand \@ifx [1]{%
 \ifx #1\expandafter \@firstoftwo
 \else \expandafter \@secondoftwo
 \fi
}%
\providecommand \natexlab [1]{#1}%
\providecommand \enquote  [1]{``#1''}%
\providecommand \bibnamefont  [1]{#1}%
\providecommand \bibfnamefont [1]{#1}%
\providecommand \citenamefont [1]{#1}%
\providecommand \href@noop [0]{\@secondoftwo}%
\providecommand \href [0]{\begingroup \@sanitize@url \@href}%
\providecommand \@href[1]{\@@startlink{#1}\@@href}%
\providecommand \@@href[1]{\endgroup#1\@@endlink}%
\providecommand \@sanitize@url [0]{\catcode `\\12\catcode `\$12\catcode `\&12\catcode `\#12\catcode `\^12\catcode `\_12\catcode `\%12\relax}%
\providecommand \@@startlink[1]{}%
\providecommand \@@endlink[0]{}%
\providecommand \url  [0]{\begingroup\@sanitize@url \@url }%
\providecommand \@url [1]{\endgroup\@href {#1}{\urlprefix }}%
\providecommand \urlprefix  [0]{URL }%
\providecommand \Eprint [0]{\href }%
\providecommand \doibase [0]{https://doi.org/}%
\providecommand \selectlanguage [0]{\@gobble}%
\providecommand \bibinfo  [0]{\@secondoftwo}%
\providecommand \bibfield  [0]{\@secondoftwo}%
\providecommand \translation [1]{[#1]}%
\providecommand \BibitemOpen [0]{}%
\providecommand \bibitemStop [0]{}%
\providecommand \bibitemNoStop [0]{.\EOS\space}%
\providecommand \EOS [0]{\spacefactor3000\relax}%
\providecommand \BibitemShut  [1]{\csname bibitem#1\endcsname}%
\let\auto@bib@innerbib\@empty
\bibitem [{\citenamefont {\ifmmode \check{Z}\else \v{Z}\fi{}uti\ifmmode~\acute{c}\else \'{c}\fi{}}\ \emph {et~al.}(2004)\citenamefont {\ifmmode \check{Z}\else \v{Z}\fi{}uti\ifmmode~\acute{c}\else \'{c}\fi{}}, \citenamefont {Fabian},\ and\ \citenamefont {Das~Sarma}}]{Spintronics_review}%
  \BibitemOpen
  \bibfield  {author} {\bibinfo {author} {\bibfnamefont {I.}~\bibnamefont {\ifmmode \check{Z}\else \v{Z}\fi{}uti\ifmmode~\acute{c}\else \'{c}\fi{}}}, \bibinfo {author} {\bibfnamefont {J.}~\bibnamefont {Fabian}},\ and\ \bibinfo {author} {\bibfnamefont {S.}~\bibnamefont {Das~Sarma}},\ }\bibfield  {title} {\bibinfo {title} {Spintronics: Fundamentals and applications},\ }\href {https://doi.org/10.1103/RevModPhys.76.323} {\bibfield  {journal} {\bibinfo  {journal} {Rev. Mod. Phys.}\ }\textbf {\bibinfo {volume} {76}},\ \bibinfo {pages} {323} (\bibinfo {year} {2004})}\BibitemShut {NoStop}%
\bibitem [{\citenamefont {Han}\ \emph {et~al.}(2018)\citenamefont {Han}, \citenamefont {Otani},\ and\ \citenamefont {Maekawa}}]{spintronic_npj_q}%
  \BibitemOpen
  \bibfield  {author} {\bibinfo {author} {\bibfnamefont {W.}~\bibnamefont {Han}}, \bibinfo {author} {\bibfnamefont {Y.}~\bibnamefont {Otani}},\ and\ \bibinfo {author} {\bibfnamefont {S.}~\bibnamefont {Maekawa}},\ }\bibfield  {title} {\bibinfo {title} {Quantum materials for spin and charge conversion},\ }\href@noop {} {\bibfield  {journal} {\bibinfo  {journal} {npj Quantum Materials}\ }\textbf {\bibinfo {volume} {3}},\ \bibinfo {pages} {27} (\bibinfo {year} {2018})}\BibitemShut {NoStop}%
\bibitem [{\citenamefont {Gentile}\ \emph {et~al.}(2017)\citenamefont {Gentile}, \citenamefont {Nacher}, \citenamefont {Saam},\ and\ \citenamefont {Walker}}]{RevModPhys_Walker_review}%
  \BibitemOpen
  \bibfield  {author} {\bibinfo {author} {\bibfnamefont {T.~R.}\ \bibnamefont {Gentile}}, \bibinfo {author} {\bibfnamefont {P.~J.}\ \bibnamefont {Nacher}}, \bibinfo {author} {\bibfnamefont {B.}~\bibnamefont {Saam}},\ and\ \bibinfo {author} {\bibfnamefont {T.~G.}\ \bibnamefont {Walker}},\ }\bibfield  {title} {\bibinfo {title} {Optically polarized $^{3}\mathrm{He}$},\ }\href {https://doi.org/10.1103/RevModPhys.89.045004} {\bibfield  {journal} {\bibinfo  {journal} {Rev. Mod. Phys.}\ }\textbf {\bibinfo {volume} {89}},\ \bibinfo {pages} {045004} (\bibinfo {year} {2017})}\BibitemShut {NoStop}%
\bibitem [{\citenamefont {Shaham}\ \emph {et~al.}(2022)\citenamefont {Shaham}, \citenamefont {Katz},\ and\ \citenamefont {Firstenberg}}]{coherent_coupling}%
  \BibitemOpen
  \bibfield  {author} {\bibinfo {author} {\bibfnamefont {R.}~\bibnamefont {Shaham}}, \bibinfo {author} {\bibfnamefont {O.}~\bibnamefont {Katz}},\ and\ \bibinfo {author} {\bibfnamefont {O.}~\bibnamefont {Firstenberg}},\ }\bibfield  {title} {\bibinfo {title} {Strong coupling of alkali-metal spins to noble-gas spins with an hour-long coherence time},\ }\href@noop {} {\bibfield  {journal} {\bibinfo  {journal} {Nat. Phys.}\ }\textbf {\bibinfo {volume} {18}},\ \bibinfo {pages} {506} (\bibinfo {year} {2022})}\BibitemShut {NoStop}%
\bibitem [{\citenamefont {Katz}\ \emph {et~al.}(2015)\citenamefont {Katz}, \citenamefont {Peleg},\ and\ \citenamefont {Firstenberg}}]{katz2015coherent_coupling}%
  \BibitemOpen
  \bibfield  {author} {\bibinfo {author} {\bibfnamefont {O.}~\bibnamefont {Katz}}, \bibinfo {author} {\bibfnamefont {O.}~\bibnamefont {Peleg}},\ and\ \bibinfo {author} {\bibfnamefont {O.}~\bibnamefont {Firstenberg}},\ }\bibfield  {title} {\bibinfo {title} {Coherent coupling of alkali atoms by random collisions},\ }\href@noop {} {\bibfield  {journal} {\bibinfo  {journal} {Phys. Rev. Lett.}\ }\textbf {\bibinfo {volume} {115}},\ \bibinfo {pages} {113003} (\bibinfo {year} {2015})}\BibitemShut {NoStop}%
\bibitem [{\citenamefont {Katz}\ \emph {et~al.}(2022)\citenamefont {Katz}, \citenamefont {Shaham},\ and\ \citenamefont {Firstenberg}}]{PRXQuantum_Firstenberg}%
  \BibitemOpen
  \bibfield  {author} {\bibinfo {author} {\bibfnamefont {O.}~\bibnamefont {Katz}}, \bibinfo {author} {\bibfnamefont {R.}~\bibnamefont {Shaham}},\ and\ \bibinfo {author} {\bibfnamefont {O.}~\bibnamefont {Firstenberg}},\ }\bibfield  {title} {\bibinfo {title} {Quantum interface for noble-gas spins based on spin-exchange collisions},\ }\href {https://doi.org/10.1103/PRXQuantum.3.010305} {\bibfield  {journal} {\bibinfo  {journal} {PRX Quantum}\ }\textbf {\bibinfo {volume} {3}},\ \bibinfo {pages} {010305} (\bibinfo {year} {2022})}\BibitemShut {NoStop}%
\bibitem [{\citenamefont {Yang}\ \emph {et~al.}(2020)\citenamefont {Yang}, \citenamefont {Liu},\ and\ \citenamefont {You}}]{yang_atom_photon_spinflip}%
  \BibitemOpen
  \bibfield  {author} {\bibinfo {author} {\bibfnamefont {F.}~\bibnamefont {Yang}}, \bibinfo {author} {\bibfnamefont {Y.-C.}\ \bibnamefont {Liu}},\ and\ \bibinfo {author} {\bibfnamefont {L.}~\bibnamefont {You}},\ }\bibfield  {title} {\bibinfo {title} {{Atom-Photon Spin-Exchange Collisions Mediated by Rydberg Dressing}},\ }\href@noop {} {\bibfield  {journal} {\bibinfo  {journal} {Phys. Rev. Lett.}\ }\textbf {\bibinfo {volume} {125}},\ \bibinfo {pages} {143601} (\bibinfo {year} {2020})}\BibitemShut {NoStop}%
\bibitem [{\citenamefont {Liu}\ \emph {et~al.}(2025)\citenamefont {Liu}, \citenamefont {Zhu}, \citenamefont {Luke}, \citenamefont {Babin}, \citenamefont {Gronowski}, \citenamefont {Ladjimi}, \citenamefont {Tomza}, \citenamefont {Bohn}, \citenamefont {Tscherbul},\ and\ \citenamefont {Ni}}]{hyperfine_to_rotational}%
  \BibitemOpen
  \bibfield  {author} {\bibinfo {author} {\bibfnamefont {Y.-X.}\ \bibnamefont {Liu}}, \bibinfo {author} {\bibfnamefont {L.}~\bibnamefont {Zhu}}, \bibinfo {author} {\bibfnamefont {J.}~\bibnamefont {Luke}}, \bibinfo {author} {\bibfnamefont {M.~C.}\ \bibnamefont {Babin}}, \bibinfo {author} {\bibfnamefont {M.}~\bibnamefont {Gronowski}}, \bibinfo {author} {\bibfnamefont {H.}~\bibnamefont {Ladjimi}}, \bibinfo {author} {\bibfnamefont {M.}~\bibnamefont {Tomza}}, \bibinfo {author} {\bibfnamefont {J.~L.}\ \bibnamefont {Bohn}}, \bibinfo {author} {\bibfnamefont {T.~V.}\ \bibnamefont {Tscherbul}},\ and\ \bibinfo {author} {\bibfnamefont {K.-K.}\ \bibnamefont {Ni}},\ }\bibfield  {title} {\bibinfo {title} {{Hyperfine-to-rotational energy transfer in ultracold atom--molecule collisions of Rb and KRb}},\ }\href@noop {} {\bibfield  {journal} {\bibinfo  {journal} {Nat. Chem.}\ }\textbf {\bibinfo {volume} {17}},\ \bibinfo {pages} {688} (\bibinfo {year} {2025})}\BibitemShut {NoStop}%
\bibitem [{\citenamefont {Chann}\ \emph {et~al.}(2002)\citenamefont {Chann}, \citenamefont {Babcock}, \citenamefont {Anderson},\ and\ \citenamefont {Walker}}]{walker_theory}%
  \BibitemOpen
  \bibfield  {author} {\bibinfo {author} {\bibfnamefont {B.}~\bibnamefont {Chann}}, \bibinfo {author} {\bibfnamefont {E.}~\bibnamefont {Babcock}}, \bibinfo {author} {\bibfnamefont {L.~W.}\ \bibnamefont {Anderson}},\ and\ \bibinfo {author} {\bibfnamefont {T.~G.}\ \bibnamefont {Walker}},\ }\bibfield  {title} {\bibinfo {title} {Measurements of ${}^{3}\mathrm{He}$ spin-exchange rates},\ }\href {https://doi.org/10.1103/PhysRevA.66.032703} {\bibfield  {journal} {\bibinfo  {journal} {Phys. Rev. A}\ }\textbf {\bibinfo {volume} {66}},\ \bibinfo {pages} {032703} (\bibinfo {year} {2002})}\BibitemShut {NoStop}%
\bibitem [{\citenamefont {Appelt}\ \emph {et~al.}(1998)\citenamefont {Appelt}, \citenamefont {Baranga}, \citenamefont {Erickson}, \citenamefont {Romalis}, \citenamefont {Young},\ and\ \citenamefont {Happer}}]{Romalis_theory}%
  \BibitemOpen
  \bibfield  {author} {\bibinfo {author} {\bibfnamefont {S.}~\bibnamefont {Appelt}}, \bibinfo {author} {\bibfnamefont {A.~B.-A.}\ \bibnamefont {Baranga}}, \bibinfo {author} {\bibfnamefont {C.~J.}\ \bibnamefont {Erickson}}, \bibinfo {author} {\bibfnamefont {M.~V.}\ \bibnamefont {Romalis}}, \bibinfo {author} {\bibfnamefont {A.~R.}\ \bibnamefont {Young}},\ and\ \bibinfo {author} {\bibfnamefont {W.}~\bibnamefont {Happer}},\ }\bibfield  {title} {\bibinfo {title} {Theory of spin-exchange optical pumping of ${}^{3}\mathrm{He}$ and ${}^{129}\mathrm{Xe}$},\ }\href {https://doi.org/10.1103/PhysRevA.58.1412} {\bibfield  {journal} {\bibinfo  {journal} {Phys. Rev. A}\ }\textbf {\bibinfo {volume} {58}},\ \bibinfo {pages} {1412} (\bibinfo {year} {1998})}\BibitemShut {NoStop}%
\bibitem [{\citenamefont {Tscherbul}\ \emph {et~al.}(2009)\citenamefont {Tscherbul}, \citenamefont {Zhang}, \citenamefont {Sadeghpour},\ and\ \citenamefont {Dalgarno}}]{main_tsherbul}%
  \BibitemOpen
  \bibfield  {author} {\bibinfo {author} {\bibfnamefont {T.~V.}\ \bibnamefont {Tscherbul}}, \bibinfo {author} {\bibfnamefont {P.}~\bibnamefont {Zhang}}, \bibinfo {author} {\bibfnamefont {H.~R.}\ \bibnamefont {Sadeghpour}},\ and\ \bibinfo {author} {\bibfnamefont {A.}~\bibnamefont {Dalgarno}},\ }\bibfield  {title} {\bibinfo {title} {Collision-induced spin exchange of alkali-metal atoms with $^{3}\text{H}\text{e}$: An ab initio study},\ }\href {https://doi.org/10.1103/PhysRevA.79.062707} {\bibfield  {journal} {\bibinfo  {journal} {Phys. Rev. A}\ }\textbf {\bibinfo {volume} {79}},\ \bibinfo {pages} {062707} (\bibinfo {year} {2009})}\BibitemShut {NoStop}%
\bibitem [{\citenamefont {Walker}(1989)}]{walker1989estimates}%
  \BibitemOpen
  \bibfield  {author} {\bibinfo {author} {\bibfnamefont {T.~G.}\ \bibnamefont {Walker}},\ }\bibfield  {title} {\bibinfo {title} {Estimates of spin-exchange parameters for alkali-metal--noble-gas pairs},\ }\href@noop {} {\bibfield  {journal} {\bibinfo  {journal} {Phys. Rev. A}\ }\textbf {\bibinfo {volume} {40}},\ \bibinfo {pages} {4959} (\bibinfo {year} {1989})}\BibitemShut {NoStop}%
\bibitem [{\citenamefont {Walter}\ \emph {et~al.}(1998)\citenamefont {Walter}, \citenamefont {Happer},\ and\ \citenamefont {Walker}}]{walter_theory}%
  \BibitemOpen
  \bibfield  {author} {\bibinfo {author} {\bibfnamefont {D.~K.}\ \bibnamefont {Walter}}, \bibinfo {author} {\bibfnamefont {W.}~\bibnamefont {Happer}},\ and\ \bibinfo {author} {\bibfnamefont {T.~G.}\ \bibnamefont {Walker}},\ }\bibfield  {title} {\bibinfo {title} {Estimates of the relative magnitudes of the isotropic and anisotropic magnetic-dipole hyperfine interactions in alkali-metal--noble-gas systems},\ }\href {https://doi.org/10.1103/PhysRevA.58.3642} {\bibfield  {journal} {\bibinfo  {journal} {Phys. Rev. A}\ }\textbf {\bibinfo {volume} {58}},\ \bibinfo {pages} {3642} (\bibinfo {year} {1998})}\BibitemShut {NoStop}%
\bibitem [{\citenamefont {Tscherbul}\ \emph {et~al.}(2011)\citenamefont {Tscherbul}, \citenamefont {Zhang}, \citenamefont {Sadeghpour},\ and\ \citenamefont {Dalgarno}}]{anisotropic_tsherbul}%
  \BibitemOpen
  \bibfield  {author} {\bibinfo {author} {\bibfnamefont {T.}~\bibnamefont {Tscherbul}}, \bibinfo {author} {\bibfnamefont {P.}~\bibnamefont {Zhang}}, \bibinfo {author} {\bibfnamefont {H.~R.}\ \bibnamefont {Sadeghpour}},\ and\ \bibinfo {author} {\bibfnamefont {A.}~\bibnamefont {Dalgarno}},\ }\bibfield  {title} {\bibinfo {title} {Anisotropic hyperfine interactions limit the efficiency of spin-exchange optical pumping of $^3${He} nuclei},\ }\href@noop {} {\bibfield  {journal} {\bibinfo  {journal} {Phys. Rev. Lett.}\ }\textbf {\bibinfo {volume} {107}},\ \bibinfo {pages} {023204} (\bibinfo {year} {2011})}\BibitemShut {NoStop}%
\bibitem [{\citenamefont {Ramsey}(1953)}]{Ramsey_main}%
  \BibitemOpen
  \bibfield  {author} {\bibinfo {author} {\bibfnamefont {N.~F.}\ \bibnamefont {Ramsey}},\ }\bibfield  {title} {\bibinfo {title} {Electron coupled interactions between nuclear spins in molecules},\ }\href {https://doi.org/10.1103/PhysRev.91.303} {\bibfield  {journal} {\bibinfo  {journal} {Phys. Rev.}\ }\textbf {\bibinfo {volume} {91}},\ \bibinfo {pages} {303} (\bibinfo {year} {1953})}\BibitemShut {NoStop}%
\bibitem [{\citenamefont {Walker}\ and\ \citenamefont {Happer}(1997)}]{walker_happer_rev}%
  \BibitemOpen
  \bibfield  {author} {\bibinfo {author} {\bibfnamefont {T.~G.}\ \bibnamefont {Walker}}\ and\ \bibinfo {author} {\bibfnamefont {W.}~\bibnamefont {Happer}},\ }\bibfield  {title} {\bibinfo {title} {Spin-exchange optical pumping of noble-gas nuclei},\ }\href@noop {} {\bibfield  {journal} {\bibinfo  {journal} {Rev. Mod. Phys.}\ }\textbf {\bibinfo {volume} {69}},\ \bibinfo {pages} {629} (\bibinfo {year} {1997})}\BibitemShut {NoStop}%
\bibitem [{\citenamefont {Marcus}(1993)}]{Marcus1993review}%
  \BibitemOpen
  \bibfield  {author} {\bibinfo {author} {\bibfnamefont {R.~A.}\ \bibnamefont {Marcus}},\ }\bibfield  {title} {\bibinfo {title} {Electron transfer reactions in chemistry. {T}heory and experiment},\ }\href {https://doi.org/10.1103/RevModPhys.65.599} {\bibfield  {journal} {\bibinfo  {journal} {Rev. Mod. Phys.}\ }\textbf {\bibinfo {volume} {65}},\ \bibinfo {pages} {599} (\bibinfo {year} {1993})}\BibitemShut {NoStop}%
\bibitem [{\citenamefont {Lorquet}\ and\ \citenamefont {Leyh-Nihant}(1988)}]{Lorquet1988NATST}%
  \BibitemOpen
  \bibfield  {author} {\bibinfo {author} {\bibfnamefont {J.~C.}\ \bibnamefont {Lorquet}}\ and\ \bibinfo {author} {\bibfnamefont {B.}~\bibnamefont {Leyh-Nihant}},\ }\bibfield  {title} {\bibinfo {title} {{Nonadiabatic unimolecular reactions. 1. A statistical formulation for the rate constants}},\ }\href@noop {} {\bibfield  {journal} {\bibinfo  {journal} {J.~Phys. Chem.}\ }\textbf {\bibinfo {volume} {92}},\ \bibinfo {pages} {4778} (\bibinfo {year} {1988})}\BibitemShut {NoStop}%
\bibitem [{\citenamefont {Lykhin}\ \emph {et~al.}(2016)\citenamefont {Lykhin}, \citenamefont {Kaliakin}, \citenamefont {dePolo}, \citenamefont {Kuzubov},\ and\ \citenamefont {Varganov}}]{Lykhin2016NATST}%
  \BibitemOpen
  \bibfield  {author} {\bibinfo {author} {\bibfnamefont {A.~O.}\ \bibnamefont {Lykhin}}, \bibinfo {author} {\bibfnamefont {D.~S.}\ \bibnamefont {Kaliakin}}, \bibinfo {author} {\bibfnamefont {G.~E.}\ \bibnamefont {dePolo}}, \bibinfo {author} {\bibfnamefont {A.~A.}\ \bibnamefont {Kuzubov}},\ and\ \bibinfo {author} {\bibfnamefont {S.~A.}\ \bibnamefont {Varganov}},\ }\bibfield  {title} {\bibinfo {title} {Nonadiabatic transition state theory: Application to intersystem crossings in the active sites of metal-sulfur proteins},\ }\href@noop {} {\bibfield  {journal} {\bibinfo  {journal} {Int. J. Quantum Chem.}\ }\textbf {\bibinfo {volume} {116}},\ \bibinfo {pages} {750} (\bibinfo {year} {2016})}\BibitemShut {NoStop}%
\bibitem [{\citenamefont {Landau}(1932)}]{Landau1932LZ}%
  \BibitemOpen
  \bibfield  {author} {\bibinfo {author} {\bibfnamefont {L.~D.}\ \bibnamefont {Landau}},\ }\bibfield  {title} {\bibinfo {title} {{Zur Theorie der Energie\"ubertragung. II}},\ }\href@noop {} {\bibfield  {journal} {\bibinfo  {journal} {Phys. Z. Sowjetunion}\ }\textbf {\bibinfo {volume} {2}},\ \bibinfo {pages} {46} (\bibinfo {year} {1932})}\BibitemShut {NoStop}%
\bibitem [{\citenamefont {Zener}(1932)}]{Zener1932LZ}%
  \BibitemOpen
  \bibfield  {author} {\bibinfo {author} {\bibfnamefont {C.}~\bibnamefont {Zener}},\ }\bibfield  {title} {\bibinfo {title} {Non-adiabatic crossing of energy levels},\ }\href {https://doi.org/10.1098/rspa.1932.0165} {\bibfield  {journal} {\bibinfo  {journal} {Proc. R. Soc. Lond. A}\ }\textbf {\bibinfo {volume} {137}},\ \bibinfo {pages} {696} (\bibinfo {year} {1932})}\BibitemShut {NoStop}%
\bibitem [{\citenamefont {Harvey}(2014)}]{Harvey2014review}%
  \BibitemOpen
  \bibfield  {author} {\bibinfo {author} {\bibfnamefont {J.~N.}\ \bibnamefont {Harvey}},\ }\bibfield  {title} {\bibinfo {title} {Spin-forbidden reactions: Computational insight into mechanisms and kinetics},\ }\href {https://doi.org/10.1002/wcms.1154} {\bibfield  {journal} {\bibinfo  {journal} {Wiley Interdiscip. Rev.: Comput. Mol. Sci.}\ }\textbf {\bibinfo {volume} {4}},\ \bibinfo {pages} {1} (\bibinfo {year} {2014})}\BibitemShut {NoStop}%
\bibitem [{\citenamefont {Miller}(1975)}]{Miller1975semiclassical}%
  \BibitemOpen
  \bibfield  {author} {\bibinfo {author} {\bibfnamefont {W.~H.}\ \bibnamefont {Miller}},\ }\bibfield  {title} {\bibinfo {title} {Semiclassical limit of quantum mechanical transition state theory for nonseparable systems},\ }\href {https://doi.org/10.1063/1.430676} {\bibfield  {journal} {\bibinfo  {journal} {J.~Chem. Phys.}\ }\textbf {\bibinfo {volume} {62}},\ \bibinfo {pages} {1899} (\bibinfo {year} {1975})}\BibitemShut {NoStop}%
\bibitem [{\citenamefont {Miller}\ \emph {et~al.}(1990)\citenamefont {Miller}, \citenamefont {Hernandez}, \citenamefont {Handy}, \citenamefont {Jayatilaka},\ and\ \citenamefont {Willetts}}]{Miller1990SCTST}%
  \BibitemOpen
  \bibfield  {author} {\bibinfo {author} {\bibfnamefont {W.~H.}\ \bibnamefont {Miller}}, \bibinfo {author} {\bibfnamefont {R.}~\bibnamefont {Hernandez}}, \bibinfo {author} {\bibfnamefont {N.~C.}\ \bibnamefont {Handy}}, \bibinfo {author} {\bibfnamefont {D.}~\bibnamefont {Jayatilaka}},\ and\ \bibinfo {author} {\bibfnamefont {A.}~\bibnamefont {Willetts}},\ }\bibfield  {title} {\bibinfo {title} {Ab initio calculation of anharmonic constants for a transition state, with application to semiclassical transition state tunneling probabilities},\ }\href {https://doi.org/10.1016/0009-2614(90)87217-F} {\bibfield  {journal} {\bibinfo  {journal} {Chem. Phys. Lett.}\ }\textbf {\bibinfo {volume} {172}},\ \bibinfo {pages} {62} (\bibinfo {year} {1990})}\BibitemShut {NoStop}%
\bibitem [{\citenamefont {Coleman}(1977)}]{Coleman1977ImF}%
  \BibitemOpen
  \bibfield  {author} {\bibinfo {author} {\bibfnamefont {S.}~\bibnamefont {Coleman}},\ }\bibfield  {title} {\bibinfo {title} {Fate of the false vacuum: {S}emiclassical theory},\ }\href {https://doi.org/10.1103/PhysRevD.15.2929} {\bibfield  {journal} {\bibinfo  {journal} {Phys. Rev. D}\ }\textbf {\bibinfo {volume} {15}},\ \bibinfo {pages} {2929} (\bibinfo {year} {1977})}\BibitemShut {NoStop}%
\bibitem [{\citenamefont {Heller}\ and\ \citenamefont {Richardson}(2020)}]{inverted}%
  \BibitemOpen
  \bibfield  {author} {\bibinfo {author} {\bibfnamefont {E.~R.}\ \bibnamefont {Heller}}\ and\ \bibinfo {author} {\bibfnamefont {J.~O.}\ \bibnamefont {Richardson}},\ }\bibfield  {title} {\bibinfo {title} {Instanton formulation of {F}ermi's golden rule in the {M}arcus inverted regime},\ }\href {https://doi.org/10.1063/1.5137823} {\bibfield  {journal} {\bibinfo  {journal} {J.~Chem. Phys.}\ }\textbf {\bibinfo {volume} {152}},\ \bibinfo {pages} {034106} (\bibinfo {year} {2020})},\ \Eprint {https://arxiv.org/abs/1911.06730} {arXiv:1911.06730 [physics.chem-ph]} \BibitemShut {NoStop}%
\bibitem [{\citenamefont {Richardson}(2024)}]{GRperspective}%
  \BibitemOpen
  \bibfield  {author} {\bibinfo {author} {\bibfnamefont {J.~O.}\ \bibnamefont {Richardson}},\ }\bibfield  {title} {\bibinfo {title} {Nonadiabatic tunneling in chemical reactions},\ }\href {https://doi.org/10.1021/acs.jpclett.4c01098} {\bibfield  {journal} {\bibinfo  {journal} {J.~Phys. Chem. Lett.}\ }\textbf {\bibinfo {volume} {15}},\ \bibinfo {pages} {7387} (\bibinfo {year} {2024})}\BibitemShut {NoStop}%
\bibitem [{\citenamefont {Dirac}(1927)}]{Dirac1927radiation}%
  \BibitemOpen
  \bibfield  {author} {\bibinfo {author} {\bibfnamefont {P.~A.~M.}\ \bibnamefont {Dirac}},\ }\bibfield  {title} {\bibinfo {title} {The quantum theory of the emission and absorption of radiation},\ }\href@noop {} {\bibfield  {journal} {\bibinfo  {journal} {Proc. R. Soc. London A.}\ }\textbf {\bibinfo {volume} {114}},\ \bibinfo {pages} {243} (\bibinfo {year} {1927})}\BibitemShut {NoStop}%
\bibitem [{\citenamefont {Wolynes}(1987)}]{Wolynes1987nonadiabatic}%
  \BibitemOpen
  \bibfield  {author} {\bibinfo {author} {\bibfnamefont {P.~G.}\ \bibnamefont {Wolynes}},\ }\bibfield  {title} {\bibinfo {title} {Imaginary time path integral {Monte Carlo} route to rate coefficients for nonadiabatic barrier crossing},\ }\href {https://doi.org/10.1063/1.453440} {\bibfield  {journal} {\bibinfo  {journal} {J.~Chem. Phys.}\ }\textbf {\bibinfo {volume} {87}},\ \bibinfo {pages} {6559} (\bibinfo {year} {1987})}\BibitemShut {NoStop}%
\bibitem [{\citenamefont {Chandler}(1998)}]{ChandlerET}%
  \BibitemOpen
  \bibfield  {author} {\bibinfo {author} {\bibfnamefont {D.}~\bibnamefont {Chandler}},\ }\bibfield  {title} {\bibinfo {title} {Electron transfer in water and other polar environments, how it happens},\ }in\ \href@noop {} {\emph {\bibinfo {booktitle} {Classical and Quantum Dynamics in Condensed Phase Simulations}}},\ \bibinfo {editor} {edited by\ \bibinfo {editor} {\bibfnamefont {B.~J.}\ \bibnamefont {Berne}}, \bibinfo {editor} {\bibfnamefont {G.}~\bibnamefont {Ciccotti}},\ and\ \bibinfo {editor} {\bibfnamefont {D.~F.}\ \bibnamefont {Coker}}}\ (\bibinfo  {publisher} {World Scientific},\ \bibinfo {address} {Singapore},\ \bibinfo {year} {1998})\ Chap.~\bibinfo {chapter} {2}, pp.\ \bibinfo {pages} {25--49}\BibitemShut {NoStop}%
\bibitem [{\citenamefont {Richardson}(2018)}]{InstReview}%
  \BibitemOpen
  \bibfield  {author} {\bibinfo {author} {\bibfnamefont {J.~O.}\ \bibnamefont {Richardson}},\ }\bibfield  {title} {\bibinfo {title} {Ring-polymer instanton theory},\ }\href {https://doi.org/10.1080/0144235X.2018.1472353} {\bibfield  {journal} {\bibinfo  {journal} {Int. Rev. Phys. Chem.}\ }\textbf {\bibinfo {volume} {37}},\ \bibinfo {pages} {171} (\bibinfo {year} {2018})}\BibitemShut {NoStop}%
\bibitem [{\citenamefont {Gutzwiller}(1990)}]{GutzwillerBook}%
  \BibitemOpen
  \bibfield  {author} {\bibinfo {author} {\bibfnamefont {M.~C.}\ \bibnamefont {Gutzwiller}},\ }\href {https://doi.org/10.1007/978-1-4612-0983-6} {\emph {\bibinfo {title} {Chaos in Classical and Quantum Mechanics}}}\ (\bibinfo  {publisher} {Springer-Verlag},\ \bibinfo {address} {New York},\ \bibinfo {year} {1990})\BibitemShut {NoStop}%
\bibitem [{\citenamefont {Miller}(1971)}]{Miller1971density}%
  \BibitemOpen
  \bibfield  {author} {\bibinfo {author} {\bibfnamefont {W.~H.}\ \bibnamefont {Miller}},\ }\bibfield  {title} {\bibinfo {title} {Classical path approximation for the {B}oltzmann density matrix},\ }\href {https://doi.org/10.1063/1.1676560} {\bibfield  {journal} {\bibinfo  {journal} {J.~Chem. Phys.}\ }\textbf {\bibinfo {volume} {55}},\ \bibinfo {pages} {3146} (\bibinfo {year} {1971})}\BibitemShut {NoStop}%
\bibitem [{\citenamefont {Feynman}\ and\ \citenamefont {Hibbs}(1965)}]{Feynman}%
  \BibitemOpen
  \bibfield  {author} {\bibinfo {author} {\bibfnamefont {R.~P.}\ \bibnamefont {Feynman}}\ and\ \bibinfo {author} {\bibfnamefont {A.~R.}\ \bibnamefont {Hibbs}},\ }\href@noop {} {\emph {\bibinfo {title} {Quantum Mechanics and Path Integrals}}}\ (\bibinfo  {publisher} {McGraw-Hill},\ \bibinfo {address} {New York},\ \bibinfo {year} {1965})\BibitemShut {NoStop}%
\bibitem [{\citenamefont {Kleinert}(2009)}]{Kleinert}%
  \BibitemOpen
  \bibfield  {author} {\bibinfo {author} {\bibfnamefont {H.}~\bibnamefont {Kleinert}},\ }\href@noop {} {\emph {\bibinfo {title} {Path Integrals in Quantum Mechanics, Statistics, Polymer Physics and Financial Markets}}},\ \bibinfo {edition} {5th}\ ed.\ (\bibinfo  {publisher} {World Scientific},\ \bibinfo {address} {Singapore},\ \bibinfo {year} {2009})\BibitemShut {NoStop}%
\bibitem [{\citenamefont {Richardson}\ \emph {et~al.}(2015)\citenamefont {Richardson}, \citenamefont {Bauer},\ and\ \citenamefont {Thoss}}]{GoldenGreens}%
  \BibitemOpen
  \bibfield  {author} {\bibinfo {author} {\bibfnamefont {J.~O.}\ \bibnamefont {Richardson}}, \bibinfo {author} {\bibfnamefont {R.}~\bibnamefont {Bauer}},\ and\ \bibinfo {author} {\bibfnamefont {M.}~\bibnamefont {Thoss}},\ }\bibfield  {title} {\bibinfo {title} {Semiclassical {G}reen's functions and an instanton formulation of electron-transfer rates in the nonadiabatic limit},\ }\href {https://doi.org/10.1063/1.4932361} {\bibfield  {journal} {\bibinfo  {journal} {J.~Chem. Phys.}\ }\textbf {\bibinfo {volume} {143}},\ \bibinfo {pages} {134115} (\bibinfo {year} {2015})},\ \Eprint {https://arxiv.org/abs/1508.04919} {arXiv:1508.04919 [physics.chem-ph]} \BibitemShut {NoStop}%
\bibitem [{\citenamefont {Fang}\ \emph {et~al.}(2023)\citenamefont {Fang}, \citenamefont {Heller},\ and\ \citenamefont {Richardson}}]{CIinst}%
  \BibitemOpen
  \bibfield  {author} {\bibinfo {author} {\bibfnamefont {W.}~\bibnamefont {Fang}}, \bibinfo {author} {\bibfnamefont {E.~R.}\ \bibnamefont {Heller}},\ and\ \bibinfo {author} {\bibfnamefont {J.~O.}\ \bibnamefont {Richardson}},\ }\bibfield  {title} {\bibinfo {title} {Competing quantum effects in heavy-atom tunnelling through conical intersections},\ }\href {https://doi.org/10.1039/D3SC03706A} {\bibfield  {journal} {\bibinfo  {journal} {Chem. Sci.}\ }\textbf {\bibinfo {volume} {14}},\ \bibinfo {pages} {10777} (\bibinfo {year} {2023})}\BibitemShut {NoStop}%
\bibitem [{\citenamefont {Ansari}\ \emph {et~al.}(2022)\citenamefont {Ansari}, \citenamefont {Heller}, \citenamefont {Trenins},\ and\ \citenamefont {Richardson}}]{PhilTransA}%
  \BibitemOpen
  \bibfield  {author} {\bibinfo {author} {\bibfnamefont {I.~M.}\ \bibnamefont {Ansari}}, \bibinfo {author} {\bibfnamefont {E.~R.}\ \bibnamefont {Heller}}, \bibinfo {author} {\bibfnamefont {G.}~\bibnamefont {Trenins}},\ and\ \bibinfo {author} {\bibfnamefont {J.~O.}\ \bibnamefont {Richardson}},\ }\bibfield  {title} {\bibinfo {title} {{Instanton theory for Fermi's golden rule and beyond}},\ }\href {https://doi.org/doi: 10.1098/rsta.2020.0378} {\bibfield  {journal} {\bibinfo  {journal} {Phil. Trans. R. Soc. A.}\ }\textbf {\bibinfo {volume} {380}},\ \bibinfo {pages} {20200378} (\bibinfo {year} {2022})}\BibitemShut {NoStop}%
\bibitem [{\citenamefont {Borel}\ \emph {et~al.}(2003)\citenamefont {Borel}, \citenamefont {S\o{}gaard}, \citenamefont {Svendsen},\ and\ \citenamefont {Andersen}}]{borel_exp}%
  \BibitemOpen
  \bibfield  {author} {\bibinfo {author} {\bibfnamefont {P.~I.}\ \bibnamefont {Borel}}, \bibinfo {author} {\bibfnamefont {L.~V.}\ \bibnamefont {S\o{}gaard}}, \bibinfo {author} {\bibfnamefont {W.~E.}\ \bibnamefont {Svendsen}},\ and\ \bibinfo {author} {\bibfnamefont {N.}~\bibnamefont {Andersen}},\ }\bibfield  {title} {\bibinfo {title} {Spin-exchange and spin-destruction rates for the ${}^{3}\mathrm{He}\ensuremath{-}\mathrm{Na}$ system},\ }\href {https://doi.org/10.1103/PhysRevA.67.062705} {\bibfield  {journal} {\bibinfo  {journal} {Phys. Rev. A}\ }\textbf {\bibinfo {volume} {67}},\ \bibinfo {pages} {062705} (\bibinfo {year} {2003})}\BibitemShut {NoStop}%
\bibitem [{\citenamefont {Ben-Amar~Baranga}\ \emph {et~al.}(1998)\citenamefont {Ben-Amar~Baranga}, \citenamefont {Appelt}, \citenamefont {Romalis}, \citenamefont {Erickson}, \citenamefont {Young}, \citenamefont {Cates},\ and\ \citenamefont {Happer}}]{Baranga_PRL}%
  \BibitemOpen
  \bibfield  {author} {\bibinfo {author} {\bibfnamefont {A.}~\bibnamefont {Ben-Amar~Baranga}}, \bibinfo {author} {\bibfnamefont {S.}~\bibnamefont {Appelt}}, \bibinfo {author} {\bibfnamefont {M.~V.}\ \bibnamefont {Romalis}}, \bibinfo {author} {\bibfnamefont {C.~J.}\ \bibnamefont {Erickson}}, \bibinfo {author} {\bibfnamefont {A.~R.}\ \bibnamefont {Young}}, \bibinfo {author} {\bibfnamefont {G.~D.}\ \bibnamefont {Cates}},\ and\ \bibinfo {author} {\bibfnamefont {W.}~\bibnamefont {Happer}},\ }\bibfield  {title} {\bibinfo {title} {Polarization of ${}^{3}\mathrm{He}$ by spin exchange with optically pumped rb and k vapors},\ }\href {https://doi.org/10.1103/PhysRevLett.80.2801} {\bibfield  {journal} {\bibinfo  {journal} {Phys. Rev. Lett.}\ }\textbf {\bibinfo {volume} {80}},\ \bibinfo {pages} {2801} (\bibinfo {year} {1998})}\BibitemShut {NoStop}%
\bibitem [{\citenamefont {Neese}(2012)}]{Neese2012orca}%
  \BibitemOpen
  \bibfield  {author} {\bibinfo {author} {\bibfnamefont {F.}~\bibnamefont {Neese}},\ }\bibfield  {title} {\bibinfo {title} {The {ORCA} program system},\ }\href {https://doi.org/10.1002/wcms.81} {\bibfield  {journal} {\bibinfo  {journal} {WIREs Comput. Mol. Sci.}\ }\textbf {\bibinfo {volume} {2}},\ \bibinfo {pages} {73} (\bibinfo {year} {2012})}\BibitemShut {NoStop}%
\bibitem [{Note1()}]{Note1}%
  \BibitemOpen
  \bibinfo {note} {Both curves were fitted to a univariate 5th-order spline to smoothen them. A smoothening factor of $10^{-11}$\protect \,a.u was applied for $V(r)$ and $10^{-15}$\protect \,a.u for $\Delta (r)$.}\BibitemShut {Stop}%
\bibitem [{\citenamefont {Fang}\ \emph {et~al.}(2021)\citenamefont {Fang}, \citenamefont {Winter},\ and\ \citenamefont {Richardson}}]{DosTMI}%
  \BibitemOpen
  \bibfield  {author} {\bibinfo {author} {\bibfnamefont {W.}~\bibnamefont {Fang}}, \bibinfo {author} {\bibfnamefont {P.}~\bibnamefont {Winter}},\ and\ \bibinfo {author} {\bibfnamefont {J.~O.}\ \bibnamefont {Richardson}},\ }\bibfield  {title} {\bibinfo {title} {Microcanonical tunneling rates from density-of-states instanton theory},\ }\href {https://doi.org/10.1021/acs.jctc.0c01118} {\bibfield  {journal} {\bibinfo  {journal} {J. Chem. Theory Comput.}\ }\textbf {\bibinfo {volume} {17}},\ \bibinfo {pages} {40} (\bibinfo {year} {2021})}\BibitemShut {NoStop}%
\bibitem [{\citenamefont {Lawrence}\ and\ \citenamefont {Richardson}(2022)}]{JoeFaraday}%
  \BibitemOpen
  \bibfield  {author} {\bibinfo {author} {\bibfnamefont {J.~E.}\ \bibnamefont {Lawrence}}\ and\ \bibinfo {author} {\bibfnamefont {J.~O.}\ \bibnamefont {Richardson}},\ }\bibfield  {title} {\bibinfo {title} {Improved microcanonical instanton theory},\ }\href {https://doi.org/10.1039/D2FD00063F} {\bibfield  {journal} {\bibinfo  {journal} {Faraday Discuss.}\ }\textbf {\bibinfo {volume} {238}},\ \bibinfo {pages} {204} (\bibinfo {year} {2022})}\BibitemShut {NoStop}%
\bibitem [{\citenamefont {Levine}(2005)}]{Levine}%
  \BibitemOpen
  \bibfield  {author} {\bibinfo {author} {\bibfnamefont {R.~D.}\ \bibnamefont {Levine}},\ }\href@noop {} {\emph {\bibinfo {title} {Molecular Reaction Dynamics}}}\ (\bibinfo  {publisher} {Cambridge University Press},\ \bibinfo {address} {Cambridge},\ \bibinfo {year} {2005})\BibitemShut {NoStop}%
\bibitem [{\citenamefont {Soboll}(1972)}]{soboll1972spin}%
  \BibitemOpen
  \bibfield  {author} {\bibinfo {author} {\bibfnamefont {H.}~\bibnamefont {Soboll}},\ }\bibfield  {title} {\bibinfo {title} {Spin exchange between optically oriented sodium and foreign gas nuclei},\ }\href@noop {} {\bibfield  {journal} {\bibinfo  {journal} {Phys. Lett. A}\ }\textbf {\bibinfo {volume} {41}},\ \bibinfo {pages} {373} (\bibinfo {year} {1972})}\BibitemShut {NoStop}%
\bibitem [{\citenamefont {Lawrence}\ \emph {et~al.}(2023)\citenamefont {Lawrence}, \citenamefont {Du{\v{s}}ek},\ and\ \citenamefont {Richardson}}]{lawrence2023perturbatively}%
  \BibitemOpen
  \bibfield  {author} {\bibinfo {author} {\bibfnamefont {J.~E.}\ \bibnamefont {Lawrence}}, \bibinfo {author} {\bibfnamefont {J.}~\bibnamefont {Du{\v{s}}ek}},\ and\ \bibinfo {author} {\bibfnamefont {J.~O.}\ \bibnamefont {Richardson}},\ }\bibfield  {title} {\bibinfo {title} {Perturbatively corrected ring-polymer instanton theory for accurate tunneling splittings},\ }\href@noop {} {\bibfield  {journal} {\bibinfo  {journal} {J.~Chem. Phys.}\ }\textbf {\bibinfo {volume} {159}},\ \bibinfo {pages} {014111} (\bibinfo {year} {2023})}\BibitemShut {NoStop}%
\bibitem [{\citenamefont {Trenins}\ and\ \citenamefont {Richardson}(2022)}]{trenins2022nonadiabatic}%
  \BibitemOpen
  \bibfield  {author} {\bibinfo {author} {\bibfnamefont {G.}~\bibnamefont {Trenins}}\ and\ \bibinfo {author} {\bibfnamefont {J.~O.}\ \bibnamefont {Richardson}},\ }\bibfield  {title} {\bibinfo {title} {Nonadiabatic instanton rate theory beyond the golden-rule limit},\ }\href@noop {} {\bibfield  {journal} {\bibinfo  {journal} {J.~Chem. Phys.}\ }\textbf {\bibinfo {volume} {156}},\ \bibinfo {pages} {174115} (\bibinfo {year} {2022})}\BibitemShut {NoStop}%
\bibitem [{\citenamefont {Tscherbul}\ \emph {et~al.}(2016)\citenamefont {Tscherbul}, \citenamefont {Brumer},\ and\ \citenamefont {Buchachenko}}]{spin_orbit_ion_atom}%
  \BibitemOpen
  \bibfield  {author} {\bibinfo {author} {\bibfnamefont {T.~V.}\ \bibnamefont {Tscherbul}}, \bibinfo {author} {\bibfnamefont {P.}~\bibnamefont {Brumer}},\ and\ \bibinfo {author} {\bibfnamefont {A.~A.}\ \bibnamefont {Buchachenko}},\ }\bibfield  {title} {\bibinfo {title} {Spin-orbit interactions and quantum spin dynamics in cold ion-atom collisions},\ }\href {https://doi.org/10.1103/PhysRevLett.117.143201} {\bibfield  {journal} {\bibinfo  {journal} {Phys. Rev. Lett.}\ }\textbf {\bibinfo {volume} {117}},\ \bibinfo {pages} {143201} (\bibinfo {year} {2016})}\BibitemShut {NoStop}%
\bibitem [{\citenamefont {Tscherbul}\ and\ \citenamefont {Krems}(2006)}]{electric_field_spin_flip}%
  \BibitemOpen
  \bibfield  {author} {\bibinfo {author} {\bibfnamefont {T.~V.}\ \bibnamefont {Tscherbul}}\ and\ \bibinfo {author} {\bibfnamefont {R.~V.}\ \bibnamefont {Krems}},\ }\bibfield  {title} {\bibinfo {title} {Controlling electronic spin relaxation of cold molecules with electric fields},\ }\href {https://doi.org/10.1103/PhysRevLett.97.083201} {\bibfield  {journal} {\bibinfo  {journal} {Phys. Rev. Lett.}\ }\textbf {\bibinfo {volume} {97}},\ \bibinfo {pages} {083201} (\bibinfo {year} {2006})}\BibitemShut {NoStop}%
\bibitem [{\citenamefont {Krems}\ and\ \citenamefont {Dalgarno}(2004)}]{krems_magnetic_depolarization}%
  \BibitemOpen
  \bibfield  {author} {\bibinfo {author} {\bibfnamefont {R.}~\bibnamefont {Krems}}\ and\ \bibinfo {author} {\bibfnamefont {A.}~\bibnamefont {Dalgarno}},\ }\bibfield  {title} {\bibinfo {title} {Quantum-mechanical theory of atom-molecule and molecular collisions in a magnetic field: Spin depolarization},\ }\href@noop {} {\bibfield  {journal} {\bibinfo  {journal} {J.~Chem. Phys.}\ }\textbf {\bibinfo {volume} {120}},\ \bibinfo {pages} {2296} (\bibinfo {year} {2004})}\BibitemShut {NoStop}%
\bibitem [{\citenamefont {Dikopoltsev}\ \emph {et~al.}(2025)\citenamefont {Dikopoltsev}, \citenamefont {Berrebi}, \citenamefont {Levy},\ and\ \citenamefont {Katz}}]{katz_decoherence}%
  \BibitemOpen
  \bibfield  {author} {\bibinfo {author} {\bibfnamefont {M.}~\bibnamefont {Dikopoltsev}}, \bibinfo {author} {\bibfnamefont {A.}~\bibnamefont {Berrebi}}, \bibinfo {author} {\bibfnamefont {U.}~\bibnamefont {Levy}},\ and\ \bibinfo {author} {\bibfnamefont {O.}~\bibnamefont {Katz}},\ }\bibfield  {title} {\bibinfo {title} {Suppressing the decoherence of alkali-metal spins at low magnetic fields},\ }\href {https://doi.org/10.1103/PhysRevLett.134.143201} {\bibfield  {journal} {\bibinfo  {journal} {Phys. Rev. Lett.}\ }\textbf {\bibinfo {volume} {134}},\ \bibinfo {pages} {143201} (\bibinfo {year} {2025})}\BibitemShut {NoStop}%
\end{thebibliography}%

\end{document}